\documentclass[showpacs,amsmath,amssymb,twocolumn,prl,superscriptaddress]{revtex4-1}

\usepackage{pifont}
\usepackage{diagbox}
\usepackage{graphicx,epsfig,subfigure,dsfont,amssymb,amsthm,amsfonts,amsbsy,mathrsfs,amscd}
\usepackage{soul}
\usepackage{epsfig}
\usepackage{times}
\usepackage{color}
\usepackage{physics}
\usepackage{framed}

\def\be{\begin{equation}}
\def\ee{\end{equation}}
\def\ba{\begin{array}}
\def\ea{\end{array}}

\usepackage{hyperref}
\hypersetup{  colorlinks=true, linkcolor=blue, citecolor=red, urlcolor=blue  }

\usepackage{picinpar,graphicx}
\begin{document}

\title{Incompatibility of Observables as State-Independent Bound of Uncertainty Relations}
%
%
%


\author{Yunlong Xiao}
\thanks{These three authors contributed equally}
\affiliation{Shenzhen Institute for Quantum Science and Engineering and Department of Physics, Southern University of Science and Technology , Shenzhen, 518055, China}
\affiliation{Institute for Quantum Science and Technology and Department of Mathematics and Statistics, University of Calgary, Calgary, Alberta T2N IN4, Canada}
\affiliation{Max Planck Institute for Mathematics in the Sciences, 04103 Leipzig, Germany}

\author{Cheng Guo}
\thanks{These three authors contributed equally}
\affiliation{Institute for Advanced Study, Tsinghua University, Beijing, 100084, China}

\author{Fei Meng}
\thanks{These three authors contributed equally}
\affiliation{Department of Computer Science, The University of Hong Kong, Pokfulam Road, Hong Kong}
\affiliation{Shenzhen Institute for Quantum Science and Engineering and Department of Physics, Southern University of Science and Technology , Shenzhen, 518055, China}

\author{Naihuan Jing}
\affiliation{Department of Mathematics, Shanghai University, Shanghai 200444, China}
\affiliation{Department of Mathematics, North Carolina State University, Raleigh, NC 27695, USA}

\author{Man-Hong Yung}
\thanks{Corresponding author: yung@sustech.edu.cn}
\affiliation{Shenzhen Institute for Quantum Science and Engineering and Department of Physics, Southern University of Science and Technology , Shenzhen, 518055, China}
\affiliation{Shenzhen Key Laboratory of Quantum Science and Engineering,Southern University of Science and Technology, Shenzhen 518055, China}
\affiliation{Central Research Institute, Huawei Technologies, Shenzhen, 518129, China}

\begin{abstract}
For a pair of observables, they are called ``incompatible", if and only if their commutator does not vanish, which represents one of the key features in quantum mechanics. The question is, how can we characterize the incompatibility among three or more observables? Here we explore one possible route towards this goal through uncertainty relations, which impose fundamental constraints on the measurement precisions for incompatible observables. Specifically, we propose to measure the incompatibility by the 
optimal state-independent bounds of additive variance-based uncertainty relations. In this way, the degree of incompatibility becomes an intrinsic property among the operators, i.e., state independent. In particular, we focus on the incompatibility of spin-$1/2$ systems as an illustration. For an arbitrary, including non-orthogonal, setting of a finite number Pauli-spin operators, the incompatibility is analytically solved; the spins are maximally incompatible if and only if they are orthogonal to each other. On the other hand, our measure of incompatibility represents a versatile tool for applications such as testing entanglement of bipartite states, and EPR-steering criteria.
\end{abstract}


\maketitle
{\bf Introduction---}As a distinguished aspect of the quantum theory, uncertainty relations (UR)~\cite{Heisenberg1927,Kennard1927,Weyl1927,Robertson1929,Schrodinger1930} represent a fundamental limitation on measurements of physical systems; it is generally impossible to simultaneously measure two complementary observables of a physical system without an ``uncertainty". Furthermore, uncertainty relations manifest many intrinsic discrepancies between classical and quantum mechanics, leading to applications such as entanglement detection \cite{Hofmann2003, Guhne2004}, nonlocality of quantum systems
\cite{Oppenheim2010}, and EPR-steering criteria \cite{Reid1989, Cavalcanti2009, Rutkowski2017, Jia2017, Xiao2018Q} etc.


One of the most well-known uncertainty relations, between a pair of observables $M_1$ and $M_2$, were formulated in terms of a commutator, $[{M_1},{M_2}] \equiv {M_1}{M_2} - {M_2}{M_1}$, by Robertson~\cite{Robertson1929} in 1929, 
\begin{align}\label{e:Robertson}
\Delta {M_1} \ \Delta {M_2} \geqslant \frac{1}{2} | \langle \psi |[{M_1},{M_2}]|\psi \rangle |  \ ,
\end{align}
where $\Delta {M_i} \equiv {(\left\langle {M_i^2} \right\rangle  - {\left\langle {{M_i}} \right\rangle ^2})^{1/2}}$ is the standard deviation for the quantum state $|\psi\rangle$. This uncertainty relation appears in almost every textbook of quantum mechanics, and is regarded as fundamental, connecting the physical concept of {\it incompatibility} of observables (IO) with quantum {\it uncertainty}.

However, Robertson's inequality cannot be regarded as complete for describing the connection between incompatibility and uncertainty. What if the state $|\psi \rangle $ is an eigenstate of $M_1$ or $M_2$? The left-hand side becomes zero, which makes no difference if $M_1$ and $M_2$ are incompatible or not. Another problem occurs when $|\psi \rangle $ is an eigenstate of the commutator associated with an eigenvalue zero, making the inequality trivial. 
These problems point to the idea that incompatibility cannot be quantified properly by uncertainty relations when they depend on quantum states~\cite{Deutsch1983}. 

To avoid such problems, Deutsch~\cite{Deutsch1983} proposed that UR should be expressed in a state-independent form:
\begin{align}\label{e:EUR}
\mathcal{U}(M_{1}, M_{2}, |\psi\rangle)\geqslant \ \mathcal{B}(M_{1}, M_{2}) \ ,
\end{align}
where the functional $\mathcal{U}$ denotes the total uncertainty, and $\mathcal{B}$ labels a \textit{tight} state-independent bound. Here $\mathcal{B}$ only depends on observables $M_{1}$, $M_{2}$ and the functional form $\mathcal{U}$, and hence it measures the intrinsic incompatibility between the two observables. State-independent URs have been investigated from the information-theoretic perspective~\cite{Partovi1983,Kraus1987,Maassen1988,Ivanovic1992,Sanchez1993,Ballester2007,Wu2009,Berta2010,Li2011,Prevedel2011,Huang2011,Tomamichel2011,Coles2012,Coles2014,Kaniewski2014, Xiao2016S, Xiao2016QM, Xiao2016U,Xiao2018H,Xiao2018Q,Coles2019}. For example, $\mathcal{U}$ can be taken to be the sum of entropies of different bases of measurements (say $\left\{ {\left| {{u_i}} \right\rangle } \right\}$ and $\left\{ {\left| {{v_j}} \right\rangle } \right\}$), and the lower bound~$\mathcal{B}$ is given by functions of the overlap of the basis vectors, $c(i,j) \equiv |\langle {u_i}|{v_j}\rangle {|^2} $. As a second example, consider a pair of Pauli spin operators ${{S_{{{\vec n}_i}}}}$ and ${{S_{{{\vec n}_j}}}}$ pointing to different directions labeled by unit vectors ${\vec n_{i,j}}$, where ${S_{{{\vec n}_{i,j}}}} \equiv {{\vec n}_{i,j}} \cdot \vec \sigma$. The optimal state-independent bound for the additive variance-based UR, ${{\Delta ^2}{S_{{{\vec n}_i}}} + {\Delta ^2}{S_{{{\vec n}_j}}}}$, is given~\cite{Busch2014} by
\begin{equation}\label{minrho2S1mnn}
\mathop {\min }\limits_\rho  \left( {{\Delta ^2}{S_{{{\vec n}_i}}} + {\Delta ^2}{S_{{{\vec n}_j}}}} \right) = 1 - |{{\vec n}_i} \cdot {{\vec n}_j}| \equiv \mathcal{I}\left( {{{\vec n}_i},{{\vec n}_j}} \right) \ .
\end{equation}
 The spin operators are compatible ($\mathcal{I}=0$), whenever ${{\vec n}_i} \cdot {{\vec n}_j} = 1$, and are maximally incompatible ($\mathcal{I}$ is maximized), whenever they are orthogonal to each other, i.e. ${{\vec n}_i} \cdot {{\vec n}_j} = 0$.

Generalizing the above observation, we propose to measure the incompatibility among three or more observables by the optimal state-independent bound $\mathcal{B}$. In order to give a proper measure, the functional form $\mathcal{U}$ should be suitably chosen. Clearly, the multiplicative variance-based $\mathcal{U}$ in Eq.\ref{e:Robertson} fails to give a proper incompatibility measure $\mathcal{B}$. Nevertheless, proper bounds can be given by the additive variance-based UR, as well as the entropic UR. As an illustration, we focus on the variance-based UR in this paper. In particular, for an arbitrary, including non-orthogonal, setting of a finite number Pauli-spin operators $\{S_{\Vec{n}_i}\}$, their incompatibility $\mathcal{I}(\{S_{\Vec{n}_i}\})$ is analytically solved in spin-$1/2$ systems. Generalizations to higher dimensional systems and for entropic UR can be made in future study.

As a measure of incompatibility, the value of $\mathcal{B}$ depends on the choice of the functional form $\mathcal{U}$. Therefore, this notion of incompatibility intimately relates to preparational uncertainty relations. In this way, it differs from the traditional incompatibility measure for a collection of measurements~\cite{Heinosaari2016}. With our notion of incompatibility measure, new criteria for entanglement and EPR-Steering detection can be given. Therefore, this new notion of incompatibility measure connects incompatibility, uncertainty relations and quantum correlations, and can be used as a versatile tool for investigating fundational questions of quantum mechanics. 

Notably, Bush \textit{et. al.} have defined the incompatibility between two measurements by the optimal state-independent bound of the Heisenberg's error-disturbance relation~\cite{Busch2014}. However, due to the intrinsic nature of error-disturbance relations, their result is limited to only two measurements. In contrast, using preparational uncertainty relations, our definition is also applicable to three or more observables.

{\bf Setting the stage---}We first focus on determining the incompatibility for an arbitrary finite set of 2-by-2 Hermitian observables, through variance-based preparational uncertainty relations. We shall later present some of the results associated with spin operators. First of all, any Hermitian operator $M_j$ can be parametrized by a number and a not necessarily normalized 3D vector, denoted by $a_j$ and ${\vec n}_j$ respectively. Explicitly, $M_j =  {a_j I + {\vec n}_j \cdot \vec \sigma }$ where $\vec{n}_j$ is not assumed to be normalized. For any given density matrix $\rho$, it can be parametrized by $\rho  = \left( {I + \vec r \cdot \vec \sigma } \right)/2$. Therefore,
\begin{equation}
{\Delta ^2}M_j = {\text{tr}}\left( {{M_j^2}\rho } \right) - {\text{t}}{{\text{r}}^2}\left( {M_j\rho } \right) = {( {{\vec n}_j \cdot {\vec n}_j} )} - {( {{\vec n}_j \cdot \vec r} )^2} \ ,
\end{equation}
which means that the variance is independent of the value of the constant $a_j$. In other words, we can instead consider the variances of a group of non-orthogonal spin operators ${S_{{{\vec n}_{j}}}} \equiv {{\vec n}_{j}}\cdot \vec{\sigma}$, i.e. 
\begin{equation}
\sum\limits_j {{\Delta ^2}{M_j}}  \quad \Leftrightarrow \quad \sum\limits_j {{\Delta ^2}{S_{{{\vec n}_j}}}} \ .
\label{normalization}
\end{equation}
This result is consistent with the notion of characterizing compatibility with a commutator, $\left[ {{M_i},{M_j}} \right] = [ {{{\vec n}_i} \cdot \vec \sigma ,{{\vec n}_j} \cdot \vec \sigma } ]$, which is also independent of the values of $a_i$ and $a_j$.

The next goal is to determine the incompatibility of the observables of non-orthogonal spins, 
\begin{equation}
\mathcal{I}({{\vec n}_1},{{\vec n}_2},..,{{\vec n}_N}) \equiv \mathop {\min }\limits_\rho  \sum\limits_{j = 1}^N {{\Delta ^2}} {S_{{{\vec n}_j}}} \ ,
\end{equation}
where the number of terms $N > 1$ is any finite integer larger than 1. For a special case of three spins $N=3$, and all spin operators are orthogonal among one another, e.g., $\left\{ {{S_x},{S_y},{S_z}} \right\}$, it is known~\cite{Hofmann2003} that 
\begin{equation}\label{SXYZg2}
\Delta^{2}S_{x}+\Delta^{2}S_{y}+\Delta^{2}S_{z}\geqslant 2 \ ,
\end{equation}
which can be saturated by any pure state of a qubit. We shall see how to recover this result as a special case.

\begin{figure}[t]
  \begin{center}
\includegraphics[width=0.35\paperwidth]{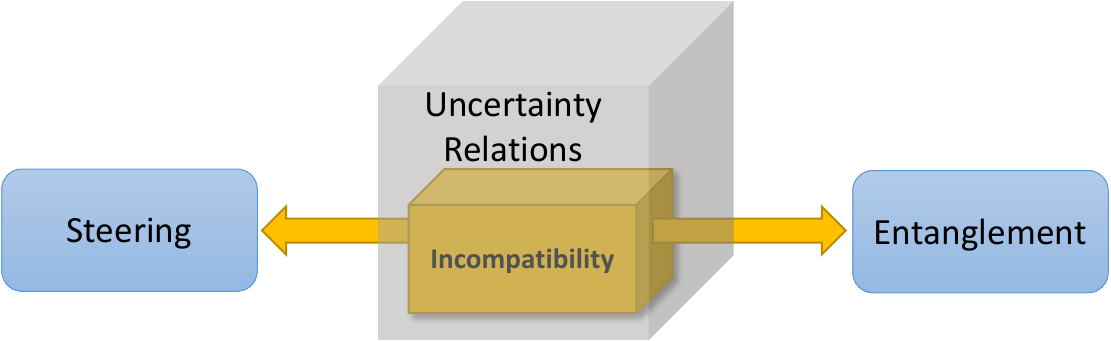}
\end{center}
\caption{The core of uncertainty relations is the incompatibility of the observables or measurement, playing the role of state-independent bound. Such a bound can be employed for detecting quantum entanglement and EPR-steering.}
\end{figure}

Let us consider again a general density matrix of a qubit, $\rho  = \left( {I + \vec r \cdot \vec \sigma } \right)/2$, where $\vec r = \left( {x,y,z} \right)$ is a normalized vector, together with ${\text{tr}}\left( {{S_{{{\vec n}_i}}}\rho } \right) = {{\vec n}_i} \cdot \vec r$ and ${\text{tr}}\left( {S_{{{\vec n}_i}}^2\rho } \right) = n_i^2$, where ${\vec n}_i$ needs not be normalized. First, in terms of the vector $\vec r$ of $\rho$, we have, 
\begin{equation}
\mathop {\min }\limits_\rho  \sum\limits_{i = 1}^N {{\Delta ^2}{S_{{{\vec n}_i}}}}  = \tau_1  - \mathop {\max }\limits_{\vec r} \sum\limits_{i = 1}^N {{{\left( {{{\vec n}_i} \cdot \vec r} \right)}^2}}  \ ,
\end{equation}
where $\tau_1 \equiv \sum\nolimits_{i=1}^N {( {{\vec n}_j \cdot {\vec n}_j} )}$ does not depend on $\vec r$. It can be shown\cite{SM} that the minimum value of the total uncertainty, over all qubit states with the constraints $\norm{\vec r}=1$, equals 
\begin{equation}
    \mathop {\min }\limits_\rho  \sum\limits_{i = 1}^N {{\Delta ^2}{S_{{{\vec n}_i}}}}  = \tau_1 - \lambda_{max}(A) \ ,
\end{equation}
where $\lambda_{max}(A) $ is the maximal eigenvalue of a 3-by-3 matrix Hermitian operator $A$ defined by
\begin{equation}
A \ \equiv \ \sum\limits_{i=1}^N \  {\left| {{{\vec n}_i}} \right\rangle \left\langle {{{\vec n}_i}} \right|} \ ,
\end{equation}
where we have adopted the Dirac notation to denote vectors with three real elements, e.g., $\left| {\vec n}_i \right\rangle  \equiv \vec{n}_i = \left( {{n_i}_x,{n_i}_y,{n_i}_z} \right)^\top$. In this way, we can write $\left\langle {{{\vec n}_i}} \right|\left. {{{\vec n}_j}} \right\rangle  = \left\langle {{{\vec n}_j}} \right|\left. {{{\vec n}_i}} \right\rangle  = {{\vec n}_i} \cdot {{\vec n}_j}$.

For example, let us first consider the special case with two non-orthogonal spins, ${\vec n}_1$, and ${\vec n}_2$, i.e. $N=2$. The eigenvalue equation is given by $\left| {{{\vec n}_1}} \right\rangle \left\langle {{{\vec n}_1}} \right|\left. {\vec r} \right\rangle  + \left| {{{\vec n}_2}} \right\rangle \left\langle {{{\vec n}_2}} \right|\left. {\vec r} \right\rangle  = \lambda \left| {\vec r} \right\rangle $, resulting in the largest eigenvalue as
\begin{equation}
{\lambda _{\max }}(A) = \frac{1}{2}\left( {{\tau _1} + \sqrt {{{\left( {n_1^2 - n_2^2} \right)}^2} + 4{{\left( {{{\vec n}_1} \cdot {{\vec n}_2}} \right)}^2}} } \right) \ ,
\end{equation}
where ${\tau _1} = {( {{\vec n}_1 \cdot {\vec n}_1} )} + {( {{\vec n}_2 \cdot {\vec n}_2} )}$. In the case of Pauli spins, where $\norm{\vec n_{1}}=\norm{\vec n_{2}}=1$, we have, ${\lambda _{\max }}\left( A \right) = 1 + \left| {{{\vec n}_1} \cdot {{\vec n}_2}} \right|$, which reduces to the result presented earlier in Eq.~(\ref{minrho2S1mnn}). 


For three or more spins, we again need to calculate the maximum eigenvalue of $A$, given by the determinant equation, $\det \left( {\lambda I - A} \right) = 0$, or explicitly, the largest root of 
\begin{equation}\label{deter_equation}
{\lambda ^3} - {\tau _1}{\lambda ^2} - \lambda \left( {{\tau _2} - \tau _1^2} \right)/2 - \det \left( A \right) = 0 \ ,
\end{equation}
where $\tau_k \equiv {\text{tr}}\left( A^k \right) $ for $k=1,2,3$. In fact, we can express the determinant, $\det A$, in terms of the $\tau$'s only, i.e., $\det A = \left( {\tau _1^3 + 2{\tau _3} - 3{\tau _1}{\tau _2}} \right)/6$.  Therefore, the characteristic equation can be completely determined by the values of the $\tau$'s. Explicitly, they are given by: (i) ${\tau _1} = \sum\nolimits_{i = 1}^N \left( {{{\vec n}_i} \cdot {{\vec n}_i}} \right) $, (ii) ${\tau _2} = {\sum\nolimits_{i,j = 1}^N {\left( {{{\vec n}_i} \cdot {{\vec n}_j}} \right)} ^2}$, and (iii) ${\tau _3} = \sum\nolimits_{i,j,k = 1}^N {\left( {{{\vec n}_i} \cdot {{\vec n}_j}} \right)} \left( {{{\vec n}_j} \cdot {{\vec n}_k}} \right)\left( {{{\vec n}_k} \cdot {{\vec n}_i}} \right)$. Therefore, we should expect that the solution of the above equation Eq.~(\ref{deter_equation}), and also the lower bound of the uncertainty relations, depends only on the products of ${{\vec n}_i} \cdot {{\vec n}_j}$, which is consistent with Eq.~(\ref{minrho2S1mnn}). 

To find the largest root, we simplify the above cubic equation (Eq.~(\ref{deter_equation})) by introducing $z \equiv \lambda -\tau_1/3$, leading to
\begin{equation}
{z^3} - 3{\alpha ^2}z - 2\beta  = 0 \ ,
\label{e:simplified_characteristic}
\end{equation}
where $\alpha  = \sqrt {\left( {3{\tau _2} - \tau _1^2} \right)/18}$, and $\beta  \equiv \tau _1^3/27 + {\tau _3}/6 - {\tau _1}{\tau _2}/6$. Since this equation is equivalently the characteristic equation of a Hermitian operator $A$, whose eigenvalues must be real numbers, we conclude that this simplified cubic equation (Eq.(\ref{e:simplified_characteristic})) must also have three real roots. In general, for a cubic equation $z^3+p z +q$ with three real roots, we have the following trigonometric solution:
\begin{equation}
    z_k = 2 \sqrt{-\frac{p}{3}} \cos \left[ \frac{1}{3} \arccos (\frac{3 q}{2 p} \sqrt{\frac{-3}{p}})-\frac{2 \pi k}{3}\right] \ ,
\end{equation}
where $k = 0, 1, 2$. Since the range of $\arccos$ is $[0,\pi]$, we can conclude that $z_0$ is the largest root among the three. Applying this to our equation, with $p=-3\alpha^2$ and $q=-2\beta$, we have  $\lambda_{max}(A) = \tau_1/3+z_0$, leading to our following main results.

  {\bf Main Results---} To illustrate that the incompatibilities among multiple observables can be measured by the optimal state-independent bound of uncertainty relations, we focus on the additive variance-based UR for qubits. By minimizing the sum of variance $\sum\limits_{j = 1}^N {{\Delta ^2}} {M_{{n_j}}}$, the incompatibility $\mathcal{I}(\{M_j\})$ is analytically calculated for an arbitrary finite set of 2-by-2 Hermitian observables $M_j = a_j I + \vec{n}_j \cdot \vec{\sigma}$, with $a_j$ being a number, $\vec \sigma$ being the Pauli matrices and $\vec{n}_j$ being an normalized 3D vector, which equals
  \begin{equation}
  \mathcal{I}\left( {\left\{ {{M_j}} \right\}} \right) \equiv \frac{{2 \tau_1}}{3} - 2\alpha \cos \left[ \frac{1}{3}{\arccos}\left( {\frac{\beta }{\alpha^{3} }} \right)\right] \ ,
  \label{eq_general_results}
\end{equation}
where $\alpha  = \sqrt {\left( {3{\tau _2} - \tau _1^2} \right)/18}$, $\beta  \equiv \tau _1^3/27 + {\tau _3}/6 - {\tau _1}{\tau _2}/6$, and $\tau$'s are defined hereinbefore. From the formula, the incompatibility among 2-by-2 Hermitian observables only depends on $\vec{n}_i \cdot \vec{n}_j$, which captures the geometric overlap between observables, and is irrelevant to the background noise $a_j$ in each observable.

Especially, if all the vectors $\vec{n}_i$ are normalized, i.e., $\norm{\vec{n}_i} = 1$ for all observables $M_j$, we have $\tau_1 = N$, ${\tau _2} = N + 2\sum\nolimits_{i < j} {{{\left( {{{\vec n}_i} \cdot {{\vec n}_j}} \right)}^2}}$, and ${\tau _3} = 3{\tau _2} - 2N + 6\sum\nolimits_{i < j < k} {\left( {{{\vec n}_i} \cdot {{\vec n}_j}} \right)\left( {{{\vec n}_j} \cdot {{\vec n}_k}} \right)\left( {{{\vec n}_k} \cdot {{\vec n}_i}} \right)}$, and the incompatibility is given by
\begin{equation}
\mathcal{I}\left( {\left\{ {{M_j}} \right\}} \right) \equiv \frac{{2N}}{3} - 2\alpha \cos (\frac{1}{3}{\cos^{ - 1}}\left( {\frac{\beta }{\alpha^{3} }} \right)),
\end{equation}
where
\begin{align}
\alpha^2=\left[\sum \limits_{1 \leqslant j < h \leqslant N} \left( \vec{n}_j \cdot
\vec{n}_h \right)^2 - N(N-3)/6 \right]/3 \ ,
\end{align}
and 
\begin{align}
\beta=&\sum_{h < j< t} \left( \vec{n}_h \cdot \vec{n}_j
\right) \left( \vec{n}_t \cdot \vec{n}_h \right) \left( \vec{n}_j \cdot
\vec{n}_t \right)\notag\\
-& \frac{1}{3} (N - 3) \sum_{j < h} \left( \vec{n}_j \cdot
\vec{n}_h \right)^2 + \frac{1}{54} N (N - 3)  (2N - 3) \ .
\end{align}
In particular, for three non-orthogonal spins ($N=3$), we have
\begin{equation}
{\alpha ^2} = \left[ {{{\left( {{{\vec n}_1} \cdot {{\vec n}_2}} \right)}^2} + {{\left( {{{\vec n}_2} \cdot {{\vec n}_3}} \right)}^2} + {{\left( {{{\vec n}_1} \cdot {{\vec n}_3}} \right)}^2}} \right]/3
\end{equation}
is the mean value of the products of ${\left( {{{\vec n}_i} \cdot {{\vec n}_j}} \right)^2}$, and 
\begin{equation}
\beta  = \left( {{{\vec n}_1} \cdot {{\vec n}_2}} \right)\left( {{{\vec n}_2} \cdot {{\vec n}_3}} \right)\left( {{{\vec n}_1} \cdot {{\vec n}_3}} \right) \ .
\end{equation}
which implies that the incompatibility, or the minimal uncertainty, for three non-orthogonal spins is given by,
\begin{equation}
\mathop {\min }\limits_\rho  \sum\limits_{j = 1}^3 {{\Delta ^2}} {S_{{n_j}}} = 2 - 2\alpha \cos \left( {\frac{1}{3} \cos{^{-1}}\left( {\frac{\beta }{\alpha^{3} }} \right)} \right) \ .
\end{equation}
This result can be reduced to the previous result Eq.~(\ref{SXYZg2}), if we choose all spin directions to be orthogonal to one another, i.e., $\left( {{{\vec n}_1},{{\vec n}_2},{{\vec n}_3}} \right) = \left( {\hat x,\hat y,\hat z} \right)$. Notably, this quantity is geometrical in the sense that it depends only on the mutual angles (inner product) between each pair of observables.

{\bf Applications---} We continue by discussing some instructive applications in entanglement and steering detection.  Firstly, the use of uncertainty arguments to study entanglement is
well known \cite{Hofmann2003}. However, their arguments are based on a unknown global minimum. With our results, the analytic expression for qubit systems can be derived. In the following, we begin with reviewing the entanglement detection via uncertainty relations in detail.

Consider incompatible observables $S_{\vec{n}_i}$, if there is no simultaneous eigenstate of
all $S_{\vec{n}_i}$, there must be a nontrivial lower limit $\mathcal{B}$ for the sum of the 
uncertainties,
\begin{align}
\sum\limits_{i}\Delta^{2}S_{\vec{n}_i}\geqslant\mathcal{B},
\end{align}
while the bound $\mathcal{B}$ is defined as the absolute minimum of the uncertainty sum for any quantum state. It therefore 
represents a universally valid limitation of the measurement statistics of quantum systems.

In general, a bipartite quantum systems between Alice and Bob can be characterized by the assemblages of incompatible observables, 
$\{S_{\vec{n}_i}\}_{i}$ and $\{S_{\vec{n}_i}\}_{j}$, with the sum uncertainty relations formulated by
\begin{align}
\sum\limits_{j}\Delta^{2}S_{\vec{n}_j}\geqslant\mathcal{B}_{j}.
\end{align}
Denote the index $j$ as the result of some permutation $\pi$, i.e. $j=\pi(i)$, then the measurement statistics of separable states 
are limited by the following uncertainty relation
\begin{align}\label{e:entanglement1}
\sum\limits_{i}\Delta^{2}(S_{\vec{n}_i}\otimes I+I\otimes S_{\vec{n}_{\pi(i)}})\geqslant\mathcal{B}_{i}+\mathcal{B}_{j},
\end{align}
which holds for all possible permutations.

To derive a experimentally feasible criterion for entanglement, $\mathcal{B}_{i}$ and $\mathcal{B}_{j}$ must have a specific expression.
Here, we can overcome this challenge easily. To show this, we consider IO on three incompatible observables. Take measurements $S_{\vec{n}_i}$ and 
$S_{\vec{m}_i}$ work on bipartite systems respectively, then for separable states, the measurement values are uncorrelated 
and the total uncertainties are limited by sum of the local uncertainties
\begin{align}\label{e:entanglement2}
&\sum\limits_{i=1}^{3}\Delta^{2}(S_{\vec{n}_i}\otimes I+I\otimes S_{\vec{m}_i})\notag\\
\geqslant \ 
&{\mathcal{I}}(\vec{n}_1, \vec{n}_2, \vec{n}_3)+
\mathcal{I}(\vec{m}_1, \vec{m}_2, \vec{m}_3).
\end{align}
Any violation of (\ref{e:entanglement2}) therefore proves that the measured quantum state cannot be separated, since entanglement describes quantum correlations that are
more precise than the ones represented by mixtures of product states. Hence the sum of the incompatibility forms a sufficient condition 
for the existence of entanglement directly.


Next, we consider the EPR-steering scenario \cite{Wiseman2007}: Alice and Bob have local access to subsystems of a 
bipartite quantum state $\rho$. Alice chooses one of her measurements $a$ with outcomes $A$, similar for Bob. Then a no-EPR-steering model for Bob is
\begin{align}
p(A, B|a, b)=\sum\limits_{\lambda}p(\lambda)p(A|a, \lambda)p_{Q}(B| b, \lambda),
\end{align}
with probability distributions $p(A|a, \lambda)$ and $p(\lambda)$ under ``hidden variable'' $\lambda$ \cite{Brunner2014}. And $p_{Q}(B| b, \lambda)$ 
represent probability distributions for outcomes $B$ which are compatible with a quantum state.

Following \cite{Cavalcanti2009}, if Alice tries to infer the outcomes of Bob's measurements through measurements on her subsystem. 
We denote by $B_{est}(A)$ Alice's estimate of the value of Bob's measurement $b$ as a function of the outcomes of her measurement $a$. 
The corresponding average inference variance of $B$ given estimate $B_{est}(A)$ is defined by
\begin{align}
\Delta^{2}_{\inf}(B)=\langle[B-B_{est}(A)]^{2}\rangle,
\end{align}
and its minimum is
\begin{align}
\Delta^{2}_{\min}(B)=\langle[B-\langle B\rangle_{A}]^{2}\rangle,
\end{align}
here the mean $\langle B\rangle_{A}$ is over the conditional probability $p(B|A)$. Under the no-EPR-steering model, we can derive a bound 
for $\Delta^{2}_{\inf}(B)$ \cite{Cavalcanti2009}
\begin{align}
\Delta^{2}_{\inf}(B)\geqslant\Delta^{2}_{\min}(B)\geqslant\sum\limits_{\lambda}p(\lambda)\Delta_{Q}^{2}(B|\lambda),
\end{align} 
where $\Delta_{Q}^{2}(B|\lambda)$ represents the probability for $B$ predicted by a quantum state $\rho_{\lambda}.$ Consequently, we can 
derive the following uncertainty relations for the no-EPR-steering model
\begin{align}
\sum_{j=1}^N \Delta^{2}_{\inf}(S_{\vec{n}_j})
\geqslant \mathcal{I}(\vec{n}_1, \cdots, \vec{n}_N).
\end{align} 
Since the above inequality follows directly from assuming no-EPR-steering model, its violation implies the non-existence of the local hidden states (LHS) model for the outcomes statistics. In other words, any violation of the above inequality works as a sufficient condition for EPR-steering. Notably, the lower bound is exactly the incompatibility $\mathcal{I}(\vec{n}_1, \cdots, \vec{n}_m)$ for Alice's observables. For qubits, applying our main results Eq.\ref{eq_general_results} gives the analytic criteria for EPR-steering.

Actually, the formalism of this criterion is based on the conditional probabilities $P(B|A)$ \cite{Cavalcanti2009}. However, 
if Alice and Bob take measurements $A_{j}$ and $S_{\vec{n}_j}$ on their own states respectively \cite{Zhen2016}, a bipartite 
state is steerable (from Alice to Bob) if the following uncertainty relations
\begin{align}
\sum_{j=1}^N \Delta^{2}(\alpha_{j}A_{j}\otimes I+ I\otimes S_{\vec{n}_j})
\geqslant \mathcal{I}(\vec{n}_1, \cdots, \vec{n}_N),
\end{align} 
is violated, and $\{\alpha_{j}\}$ are arbitrary real numbers. In all of the above we have shown the strength of steerability 
is determined by the strength of preparation uncertainty in measurements, i.e. incompatibility. The concepts of uncertainty,
entanglement and EPR-Steering are linked through the incompatibility $\mathcal{I}(\vec{n}_1, \cdots, \vec{n}_N)$.  

{\bf Conclusion---} In this Letter, we quantify the incompatibility of three or more observables by the optimal state-independent bound of uncertainty relations. As illustration,  we explicitly calculate the incompatibility of any set of 2-by-2 observables. Future investigations can be made to extend our results to higher dimensions as well as quantifying the incompatibility using various forms of uncertainty relations, such as entropic uncertainty relations and weighted uncertainty relations. It is also an interesting open problem to figure out the relationship between our notion of incompatibility measure and many other definitions \cite{Heinosaari2016,Busch2007}.

Our work 
established intriguing connections among a number of fascinating subjects, including quantum foundations, uncertainty principle, quantum 
correlations and the geometry of quantum state space, which are of interest to researchers from diverse fields. Uncertainty relations are nothing but mathematical manifestation of the incompatibility of observables, 
and that is why both entropic uncertainty relations and variance-based uncertainty relations can be used to detect entanglement and characterize steering. Note that compare with previous developments on approximating the optimal bound ~\cite{Schwonnek2017} , our method provides an analytically expression of the optimal bound for qubit states.  


\smallskip
\noindent{\bf Acknowledgments}\, We thank S. Cheng, R. Schwonnek, L. Dammeier and J. Kaniewski for correspondence. This work is supported by the National Natural Science Foundation of China (No. 11875160), the NSFC Guangdong Joint Fund (U1801661), the Guangdong Innovative and Entrepreneurial Research Team Program (No. 2016ZT06D348), Natural Science Foundation of Guangdong Province (2017B030308003), the Science, Technology and Innovation Commission of Shenzhen Municipality (JCYJ20170412152620376, JCYJ20170817105046702, ZDSYS201703031659262), and the Simons Foundation (198129).


\end{document}